\documentstyle[preprint,aps,epsf]{revtex}
\global\firstfigfalse
\global\firsttabfalse
\begin{document}
\bibliographystyle {prsty}

\def\be{\begin{equation}}
\def\ee{\end{equation}}
\def\la{\langle}
\def\ra{\rangle}

\title{
Differences between Statistical Mechanics and Thermodynamics on
the Mesoscopic Scale
}
\author{Alex Kamenev and Yuval Gefen }
\address{
Department of Condensed Matter Physics, The Weizmann Institute of Science,
Rehovot 76100, Israel. 
}

\maketitle

\begin{abstract}

We present a systematic expansion in the ratio between the level spacing and
temperature and employ it to evaluate differences between statistical
mechanics and thermodynamics in finite disordered systems. These differences
are related to spectral correlations in those systems. They are fairly
robust and are suppressed at temperatures much higher than the level spacing.

\end{abstract}
\pacs{PACS numbers: 71.10.Ca, 73.23-b, 73.23.Ps  }


One of the major results that emerged from the work in the field of
mesoscopic physics was the understanding that physical quantities may vary
{\it qualitatively} depending on the averaging procedure employed, with a
particular emphasis on differences between the canonical (CE) and the
grandcanonical (GCE) ensembles. This observation became particularly
apparent in the study of persistent currents in normal rings 
\cite{Cheung88,Imry90}  and has
been extended to other thermodynamic  
\cite{Zuzin92,Hajdu93,Stone95} and transport 
\cite{Shklovskii82,Kamenev94,Engelhardt96} 
properties. A peculiarity of mesoscopic systems is the existence of an energy 
$E_c$ associated with spectral correlations on scales  larger 
than the mean level spacing, $\Delta$. Earlier works 
\cite{Denton73,Halperin86,Nemeth90,Cheung94} 
have addressed differences between the 
ensembles at energies (or temperatures) comparable with $\Delta$. Here we shall 
study differences between the CE and GCE which persist up to energy scales 
much larger than $\Delta$. 
The term ``canonical"  deserves elaboration. One considers 
an ensemble of finite systems (typically conducting), which are {\it
macroscopically} similar but differ from each other by their impurity
configurations, details of the boundaries etc. We now assign $N^{(i)}$
electrons to the $i^{th}$ member of the ensemble; $N^{(i)}$ is selected
arbitrarily, and is uncorrelated with the energy spectrum of the system.
Moreover, it is unchanged as we vary some external parameter, $x$
(representing e.g., an Aharonov-Bohm flux). In particular one may select
$N^{(i)}=N=\mbox{const}$ for all $i$.

\par 
The idea introduced in Refs. \cite{Imry90} was to represent each member of 
the
canonical ensemble as a grandcanonical system with an {\it effective}
chemical potential, $\mu^{(i)}(x)$, which is sample specific and depends on
the parameter $x$. A canonically averaged quantity can be obtained by
expanding around a ``wrong", average, chemical potential, $\mu$ (which is
independent of $i$ and $x$), and then calculating the expansion terms
(taken at $\mu$, i.e., performed grandcanonically). This procedure, leading
to remarkable differences between CE and GCE averages, is clearly warranted
at zero temperature, T=0. In that case differences between the CE and the
GCE are solely due to averaging over quenched disorder: each system
individually can be equally described by either $N$ or $\mu^{(i)}(x)$, such
that \cite{foot1} 
\be 
\label{eq1}
-{\partial\Omega^{(i)}\over\partial\mu}\bigg|_{\mu^{(i)}(x)}\equiv 
N\left(\mu^i(x)\right)=N\, .
\ee
Here $\Omega^{(i)}(\mu,x)$ is the (sample specific) thermodynamic (TD) 
potential. The procedure alluded to
above has also been employed at $T>0$, defining the sample specific
$\mu^i(x)$ through Eq.~(\ref{eq1}). 
In that case a canonical system characterized
by a fixed number of particles, $N$, is replaced by a grandcanonical system
whose expectation value of the number of electrons is $N$ (at any given value 
of $x$). According to this picture differences between
the CE and the GCE are again due to quenched disorder (but may depend on
temperature). The starting point here is the TD relation
\be 
\label{eq2} 
F^{(i)}(N,x) = \Omega^{(i)}(\mu,x)+ \mu  N
\ee
($F^{(i)}$ is the free energy), leading to the identity:
\be 
\label{eq3}
{\partial F^{(i)}\over\partial x}\bigg|_N = 
{\partial\Omega^{(i)}\over\partial x}\bigg|_{\mu^{(i)}(x)} \, ,
\ee
where $N$ is related to $\mu^{(i)}(x)$ through Eq.~(1). 
We should recall, though,
that the TD relations, Eqs.~(\ref{eq2}) and (\ref{eq3}), are only approximate 
(see Eq.~(\ref{eq6}) below) when it
comes to statistical mechanics (SM). It is the latter which should be 
employed to calculate expectation values of equilibrium observables. 
$F^{(i)}$ should be derived microscopically, and {\em not} by calculating 
$\Omega^{(i)}$ microscopically and then employing 
Eqs.~(\ref{eq1}), (\ref{eq2}). 
To understand what is missing in the procedure described above, one may take 
as an example a system of non--interacting electrons: the occupation 
probability for a canonical system is {\em not} given by the Fermi--Dirac 
function with an effective $\mu$ \cite{Stone95,Denton73,Nemeth90,Cheung94}. 
Differences between canonically
and grandcanonically averaged quantities should reflect two elements: 
(i) sample--to--sample fluctuations due to quenched disorder; 
(ii) the deviation of SM from TD resulting from the fact that the probability 
to find the system in a given quantum state differs between the GCE and the 
CE. The latter element, manifested in corrections to Eqs. (\ref{eq2}) and 
(\ref{eq3}), occurs in finite systems \cite{foot2} and  only at finite T. 

\par 
The purpose of the present work is to elucidate some basic questions
concerning the thermodynamics of finite systems. We report on a systematic
study of differences between the CE and the GCE, yielding both the
contributions due to quenched disorder and due to differences between SM
and TD. We specifically consider non--interacting electrons moving in a 
diffusive disorder at $T>\Delta$. We
demonstrate our approach by evaluating both the persistent current and the
heat capacity. We indicate how both types of contributions reflect spectral
correlations in the system. Our analysis also points out how the 
thermodynamic limit (where differences between SM and TD vanish) is
approached. Our general scheme should allow for the study of other regimes
of disorder (e.g. the dirty ballistic), other types of averaging (e.g. over
energy), and the inclusion of electron--electron interactions.

\par 
The canonical partition function is given by
\be
\label{eq4}
Z^{(i)}(N)=\mbox{Tr}\left\{\delta(\hat N- N)e^{-\beta\hat H^{(i)}}\right\}= 
{1\over 2\pi T}
\int\limits^{i\pi T}_{-i\pi T} d\mu\; e^{-\beta(\Omega^{(i)}(\mu)+\mu N)}\, , 
\ee
where hatted quantities are operators, 
$e^{-\beta\Omega}\equiv \mbox{Tr}\{e^{-\beta(\hat H-\mu\hat N)} \}$ 
and $\beta=1/k_B T$. Since the 
eigenvalues of $\hat N$ are integers, the $\delta$-function is in fact a
Kronecker function. We note that evaluation of the integral in 
Eq.~(\ref{eq4}) 
within saddle point approximation \cite{Rosenfeld61,Kubo75}, leads to
the TD relation Eq.~(\ref{eq2}). The saddle point $\bar\mu^{(i)} (x)$ 
is sample specific. 
Instead we expand around a constant $\bar\mu$, defined by $\la N^{(i)}
(\bar\mu)\ra=N$, where $\la\ldots\ra$ denotes averaging over 
realizations of the disorder. Defining \cite{foot3}  
$\mu-\bar\mu\equiv i\sqrt{\Delta T}\, \tau$ and 
$V^{(i)}_n\equiv (n! T)^{-1} (-\Delta T)^{n/2}  
\partial^{n-1} \delta N^{(i)}/\partial\bar\mu^{n-1} $ 
we obtain 
\be 
\label{eq5} 
Z^{(i)}(N)=\sqrt{\Delta\over 2\pi T}\,\,  
e^{-\beta(\Omega^{(i)}(\bar\mu)+\bar\mu N)}
\!\!  \int\limits^{\pi\sqrt{T/\Delta}}_{-\pi\sqrt{T/\Delta}}
\!\!\! {d\tau \over\sqrt{2\pi}}\, e^{-{\tau^2/2}}\exp \left[\sum^\infty_{n=1}
V^{(i)}_n\tau^n\right]\, .
\ee
Here 
$\delta N^{(i)}(\bar\mu,x)
\equiv -\partial\Omega^{(i)}/\partial\mu\bigg|_{\bar\mu}-N$ 
is the {\it sample specific} deviation from the mean number of electrons. 
For $\Delta/T\ll 1$ we may replace the integration limits in 
Eq.~(\ref{eq5}) by $\pm\infty$.
Expanding the term $\exp[\sum_{n} V^{(i)}_n \tau^n]$, we obtain a
diagrammatic expansion where the $\{V^{(i)}_n\}$ play the role of $n-th$ order
vertices (see below). We stress that the expansion around $\bar\mu$ is {\it
not} a standard one: $\bar\mu$ does not represent the sample specific
saddle point. This expansion is justified {\it a-posteriori} (see below).
Employing linked cluster diagrammatic expansion \cite{Negele}, 
the free energy may be written
\be 
\label{eq6} 
F^{(i)}(N)=(\Omega^{(i)}(\bar\mu)+\bar\mu N)+{T\over 2}
\ln \left({2\pi T\over\Delta}\right) - 
\{\mbox{all}\,\, connected\,\, \mbox{diagrams} \}^{(i)}\, .
\ee 
The first term in Eq.~(\ref{eq6}) yields sample specific grandcanonical
observables. The second term due to Gaussian fluctuations around the
expansion point at $\bar\mu$, is a T--dependent sample--independent
contribution. 

\par
\begin{figure}[htbp]
\epsfysize=4cm
\begin{center}
\leavevmode
\epsfbox{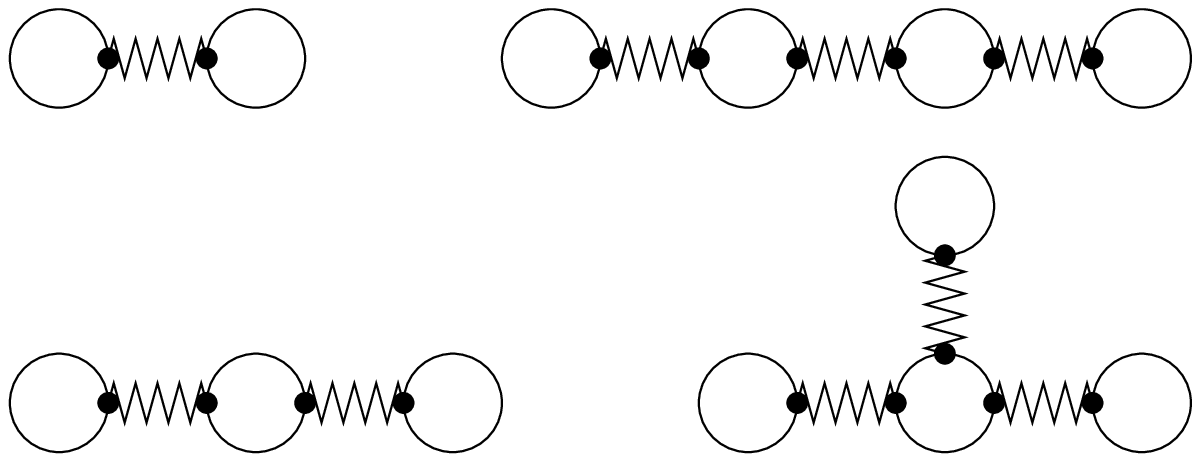}
\end{center}
\caption{\label{f1}  
The $w=0$ family of skeleton canonical diagrams with $k=1,2,3$. 
Zigzag lines are ''statistical 
lines''; full lines -- electron propagators; black dots -- scalar vertices. 
The first diagram upon averaging over disorder yields Altshuler--Shklovskii 
result. 
}
\end{figure}

\begin{figure}[htbp]
\epsfysize=4cm
\begin{center}
\leavevmode
\epsfbox{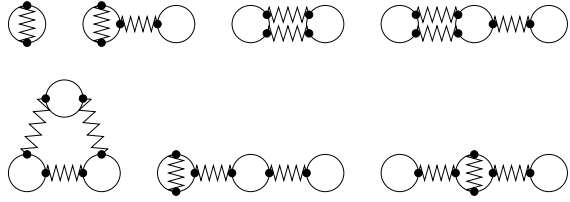}
\end{center}
\caption{\label{f2}  
Skeleton diagrams of the $w=1$ family (up to $k=3$). 
}
\end{figure}

 
We next consider the diagrammatic expansion of $F^{(i)}(N)$. 
This expansion involves integrals over $\tau$. 
In a somewhat artificial, but convenient, analogy with standard
perturbation theory, we refer to the term 
$(2\pi)^{-1/2} \int\!d \tau \exp\{-\tau^2/2\}\tau^2=1$ as a contraction or a 
``free propagator'', symbolically
$\la{\tau\tau}\ra$. Similarly a factor 
$(2\pi)^{-1/2}\int\! d\tau \exp\{-\tau^2/2\}\tau^4=3$ 
involves 3 different ways of contraction, $\la \tau\tau\tau\tau\ra =3$. 
Our ``propagators", or {\em statistical lines}, carry neither energy nor
momentum, and are employed to identify contributing diagrams and to evaluate
combinational factors. Each propagator is then represented by a ``statistical
line"  and each factor $V_n$ corresponds to an
$n-th$ order vertex, representing  an  electron loop with $n$ 
scalar vertices. Examples of  connected diagrams contributing to
$F^{(i)}(N)$ are shown in Figs. \ref{f1}, \ref{f2}. 
These are   skeleton diagrams which are yet to be
dressed by impurity lines, interaction  lines, etc. 

\par 
The following set of rules applies for the calculation of the skeleton
diagrams: (1) consider a connected diagram consisting of $p$ loops with
$n_1,\ldots,n_p$ scalar vertices respectively. We require
$\sum^p_{i=1}n_i=2k$, with $k$ being the number of statistical lines,
$k\ge p-1$. (2) Each statistical line carries a factor $(-\Delta)$. (3)
Each $n$-vertex loop carries a factor ${1\over n!}
{\partial^{n-1}\over\partial\bar\mu^{n-1}}\delta N\bigg|_{\bar\mu}\, .$ 
(4) Each diagram carries a factor $T^{w}$, where $w\equiv k-p+1 \geq 0$. 
(5) A factor of $1/m!$ should be assigned to any subset of $m$ loops
having the same number of vertices. (6) Each diagram should be multiplied 
by a combinational factor reflecting the number of  different ways to 
interconnect vertices by lines, which result in the very same diagram.

\par 
We now divide the skeleton diagrams into families, characterized by
the index $w$. Examples are depicted in Figs. \ref{f1}, \ref{f2}. It is 
understood that 
disorder averaging has to be carried out subsequently. 
Let us first consider the $w=0$ family, Fig. \ref{f1}. 
It turns out that these are
the contributions (describing differences between the CE and the GCE),
obtained when Eqs.~(\ref{eq2}) and (\ref{eq3}) are assumed to hold and each 
member of the
canonical ensemble is assigned an effective sample specific chemical
potential: the $w=0$ family represents contributions due to quenched
disorder, but {\it not} due to differences between SM and TD. (It is the only
contribution which survives at $T=0$). After some algebra one obtains 
(hereafter we suppress index $(i)$ and consider only ensemble averaged 
quantities) 
\be 
\label{eq7} 
\langle \delta F_{w=0} \rangle = - 
\sum^{\infty}_{k=1} {(-\Delta)^{k}\over (k+1)!}\, 
{\partial^{k-1}\over\partial\bar\mu^{k-1} } \langle \delta N^{k+1} \rangle \, .
\ee
Exactly the same result may be obtained by the direct solution of 
Eqs.~(\ref{eq1}) and (\ref{eq2}) \cite{Kamenev95}. 
The first ($k=1$) term in the sum  corresponds to the two loop diagram in 
Fig.~\ref{f1}. Upon averaging over (diffusive) disorder, it yields 
the Altshuler--Shklovskii term \cite{Altshuler86}, which has been employed in 
Ref. \cite{Imry90}. The
$k\ge 2$ terms are given by  complete derivatives with respect to
$\bar\mu$, and are negligible upon averaging (being small in the parameter 
$\Delta/\bar\mu$).  
This provides us with an {\em a--posteriori} justification of the 
diagrammatic expansion: sample specific terms in this expansion are not 
necessarily small, but the ensemble average is well--behaved. 

We next include the 
$w\ge 1$ families too. These yield contributions due to deviations of
SM from TD. It can be shown that for the regime where we employ our
expansion, $T>\Delta$, the leading term (in $\Delta/T$) of each family
is represented by a two-loop diagrams (cf. Figs. \ref{f1}, \ref{f2}). 
Evaluation of these diagrams  (neglecting full derivatives with 
respect to $\bar\mu$) results in 
\be 
\label{eq8}
\left\la\delta F_{two-loop}\right\ra = {\Delta\over 2} \sum^\infty_{w=0}
{(\Delta  T)^w \over (w+1)!} \left\la\left({\partial^w \delta N
\over\partial \bar\mu^w}\right)^2\right\ra \, .
\ee
The $w=0$ term is the Altshuler--Shklovskii quenched disorder contribution
\cite{Altshuler86}; $w$=1 is the leading SM vs. TD contribution \cite{foot4}.

\par 
To evaluate these terms  we define the correlator $K$ and its Fourier
transform $\tilde K$:
\be
\label{eq9}
K(\epsilon-\epsilon') \equiv 
\Delta^2 \left( \la\nu(\epsilon)\nu(\epsilon')\ra - 
\la\nu(\epsilon)\ra\la\nu (\epsilon') \ra \right) \equiv 
{1\over 2\pi} \int\limits^{+\infty}_{-\infty} dt 
\tilde K(t) e^{it(\epsilon-\epsilon')/\Delta }\, ,
\ee
where $\nu(\epsilon)$ is the sample specific density of states and $t$ is the 
dimensionless time (in units of $\hbar/\Delta$). We can write 
\be
\label{eq10}
\left\la \left({\partial^w\delta N\over\partial\bar\mu^w}\right)^2\right\ra =
{\partial^w\over\partial\bar\mu^w}{\partial^w\over\partial\bar\mu'^w} 
\int\!\!\!\!\int\limits^{+\infty}_{-\infty}
\frac{d\epsilon\,d\epsilon'}{\Delta^2} 
f(\epsilon-\bar\mu) f(\epsilon'-\bar\mu') K(\epsilon-\epsilon')
\bigg|_{\mu'=\mu} \, , 
\ee
where $f$ is the Fermi-Dirac function.  Changing variables to
$\xi\equiv\epsilon-\epsilon', \;\eta\equiv (\epsilon+\epsilon')/2$, 
performing the
integral over $\eta$, and Fourier transforming with respect to $\xi$, we
obtain
\be 
\label{eq11} 
\la\delta F_{two-loop}\ra = {\pi T\over 2} \sum^\infty_{w=0} {1\over (w+1)!}
\left({\Delta\over T}\right)^w    \int\limits^\infty_0\! dt\, 
{t^{2w} \over \mbox{sinh}^2\pi t}\, \tilde K\left( t{\Delta\over T}\right)\, . 
\ee
Eqs.~(\ref{eq6}) and (\ref{eq11}) form the basis for the analysis of the 
various corrections to the GCE
averages, and depict the dependence of these corrections on spectral
correlations. We 
study two examples comparing the $w$=0 (quenched disorder
contribution) and
$w$=1 (leading contribution due to differences between SM and TD).

The average {\it persistent current} in the canonical ensemble, 
$\la I\ra_{CE}$, is obtained by deriving $\la F\ra$ with respect to the 
Aharonov-Bohm flux, $\phi$.  The flux dependent part of the time correlator 
for  quasi--one--dimensional rings is given by 
\be 
\label{eq12} 
\tilde K (t)={|t| \over \pi} \sqrt{1\over 4 \pi g |t|}\, \sum^\infty_{p=1}
\, e^{-{p^2\over 4g|t|}} \cos4\pi p\phi/\phi_0\, , 
\ee 
where $g=E_c/\Delta\gg 1$ is the dimensionless conductance ($E_c$ is the 
Thouless correlation energy) and $\phi_0=hc/e$. 
Only even harmonics appear in Eq.~(\ref{eq12}). 
Expanding $\la I\ra_{CE}=\sum^{\infty}_{p=1} I_p \sin4\pi p\phi/\phi_0$, 
we find that the quenched disorder contribution is \cite{Imry90} 
\be 
\label{eq13} 
I_p^{dis}= {\Delta\over\phi_0}\cases{ 
{\displaystyle {2\over\pi} } 
                              \quad & $\Delta<T<E_c/p^2$\cr
{\displaystyle p^2{T\over E_c} e^{-\sqrt{2\pi T\over E_c}\,p} } 
                             \quad & $T>E_c/p^2$\,\,  ,  \cr} 
\ee
while the leading SM vs. TD correction is given by 
\be 
\label{eq14}
I_p^{SM-TD}={\Delta\over\phi_0}{1\over g} \cases{
{\displaystyle 
{15\sqrt{2\pi}\zeta(5/2)\over 64\pi^3}\, p\, \sqrt{E_c\over T}\,\, }
                              \quad & $\Delta<T<E_c/p^2$\cr
{\displaystyle 
{1\over 16\pi }\, p^4 {T\over E_c} e^{-\sqrt{2\pi T\over E_c}\, p} }
                              \quad & $T>E_c/p^2$\, .  \cr} \ee 
The dependence of the SM--TD contribution on temperature and flux is
significantly
different from the contribution due to quenched disorder (Eq. (\ref{eq13})), 
which
has been found previously. The former shows that differences between SM
and TD decrease slowly (algebraically) with temperature up to 
$T\sim E_c\gg\Delta$, and then are suppressed exponentially.

\par 
Average canonical {\it heat capacity} $(\la C\ra_{CE})$. It is given by 
$\la C\ra_{CE}= -\beta^2 \partial^2(\beta\la F\ra)/\partial \beta^2$ 
and is written as a sum of contributions (cf. Eq. (\ref{eq6})): 
\be
\label{eq15}
\la C\ra_{CE} =\la C\ra_{GCE} +\la \delta C^{Gauss}\ra +
\la\delta C^{dis}\ra + \la \delta C^{SM-TD}\ra \, . 
\ee 
The first term is the grandcanonical contribution, which for a degenerate
gas of non-interacting electrons is ${\pi^2\over 3}{T\over\Delta}$. The
term due to Gaussian fluctuations yields $-{1\over 2}$. This contribution
can be reinterpreted as a shift of the grandcanonical temperature
$(T>\Delta)$ towards a lower temperature, 
$T\rightarrow T-{3\over 2\pi^2} \Delta$ forced by the canonical constraints. 
A similar shift has been found in Ref.~\cite{Denton73}. 
The next terms in Eq.~(\ref{eq15}) are evaluated employing Eq.~(\ref{eq11}). 
We are interested in energies larger than
$\Delta$, hence $t<1$. The time correlator has two interesting regimes
(random matrix theory and Altshuler-Shklovskii \cite{Altshuler86} 
respectively): 
\be  
\label{eq16}
\tilde K(t) = {|t| \over b \pi } \cases
{1                   \quad &$g^{-1}<t<1$\cr
(4 \pi g|t|)^{-d/2}                   
                     \quad & $\Delta \tau_{e\ell} <t < g^{-1} $\cr} \, ,
\ee        
where $b=1,2,4$ for the orthogonal, unitary and symplectic ensembles
respectively. Here $\tau_{e\ell}$ is the elastic mean free time. 
Eq.~(\ref{eq16}) does not account for the crossover regimes.  
This leads  to  
\be 
\label{eq17}
\la \delta C^{dis}\ra ={1\over b \pi}  {\Delta\over T} \cases{
{\displaystyle 
{1\over 2\pi} }                \quad &   $\Delta<T<E_c$\cr
{\displaystyle 
\gamma_d \left( {T\over E_c} \right)^{d/2}   } 
                               \quad & $T>E_c$\cr}
\ee
and in the leading order in $\Delta/T$ \cite{foot5} 
\be 
\label{eq18}
\la \delta C^{SM-TD}\ra =
             {1\over b \pi} \left({\Delta\over T}\right)^2 \cases{
{\displaystyle 
{3\zeta(3)\over 4\pi}  }
                               \quad &    $\Delta <T<E_c$\cr 
{\displaystyle 
\eta_d  \left( {T\over E_c} \right)^{d/2}  }
                               \quad & $T>E_c$\, .   \cr}
\ee
In this example the SM--TD contribution is parametrically smaller than the
contribution due to  quenched disorder. 
Here too, differences between SM and TD decay
slowly over scales larger than $\Delta$: as a power law, at least over 
energy up to $\hbar/\tau_{el}$. 
The results depicted in Eqs.~(\ref{eq17}) and (\ref{eq18}), as well as an 
evaluation of $\la \delta C^{SM-TD}\ra$ for a Poissonian spectrum 
\cite{foot6} lead us to conclude that the {\it less} rigid the
spectrum the slower the decay with temperature.  

\par 
In summary, we have presented a systematic expansion in $\Delta/T$
describing differences between statistical mechanics and thermodynamics of
finite systems. 
These differences, related to spectral correlations, are found to be fairly
robust and slowly suppressed with $T$. The thermodynamic limit where they are
suppressed is attained at temperatures much higher than the level spacing.

\par 
We have benefited from discussions with Y.~Imry, A.~Schmid and A.~D.~Stone. 
This research was supported by the German--Israel 
Foundation (GIF), the U.S.--Israel Binational Science Foundation (BSF) 
and the Israel Academy of Sciences.



\begin{references}

\bibitem{Cheung88}H. F. Cheung, Y. Gefen, E. K. Riedel, and W. H. Shih, 
Phys. Rev. {\bf B 37}, 6050 (1988); 
H. Bouchiat, and G. Montambaux, J. Phys. (Paris) {\bf 50}, 2695 (1989). 

\bibitem{Imry90}Y. Imry, in\ {\em Quantum Coherence in Mesoscopic Systems},
edited by B. Kramer, NATO ASI series Vol. 254, 221--236; 
A. Schmid,  Phys. Rev. Lett. {\bf 66}, 80 (1991); 
B. L. Altshuler, Y. Gefen, and Y. Imry, Phys. Rev. Lett. {\bf 66}, 88 (1991); 
F. von Oppen, and E.K. Riedel, Phys. Rev. Lett. {\bf 66}, 84 (1991); 
E. Akkermans, Europhys. Lett {\bf 15}, 709 (1991); 
A.~Kamenev, and Y.~Gefen, Phys. Rev. Lett., {\bf 70}, 1976 (1993).

\bibitem{Zuzin92}R.~A.~Serota, and A.~Yu.~Zyuzin, Phys. Rev. {\bf B 47}, 
6399 (1993); B. L. Altshuler, Y. Gefen, Y. Imry, and G.~Montambaux, 
Phys. Rev. {\bf B 47}, 10335 (1993).

\bibitem{Hajdu93}M.~Jansen, U.~Gummich, J.~Hajdu, O.~Viehweger, Annal. 
Physik, {\bf 2}, 361 (1993). 

\bibitem{Stone95}H.~Mathur, M.~G\"ok\c cedac\u g, and A.~D.~Stone,  
Phys. Rev. Lett., {\bf  74}, 1855 (1995). 


\bibitem{Shklovskii82}B. I. Shklovskii, Pis'ma Zh. Eksp. Teor.
Fiz. {\bf 36}, 287 (1982) [JETP Lett. {\bf 36}, 352 (1982)].

\bibitem{Kamenev94} A.~Kamenev, B.~Reulet, H.~Bouchiat, and Y.~Gefen, 
Europhys. Lett., {\bf 28}, 391--396, (1994); W.~Lehle, and A. Schmid,
Annal. Physik, {\bf 4}, 451 (1995). 

 
\bibitem{Engelhardt96}M.~Engelhardt, Phys. Rev., {\bf D 52}, 1267 (1995).

\bibitem{Denton73}R.~Denton, B.~M\"uhlschlegel, and D.~J.~Scalapino, 
Phys. Rev. {\bf B 7}, 3589 (1973). 

\bibitem{Halperin86}W.~P.~Halperin, Rev. of Modern. Phys., {\bf 58}, 533 
(1986); J. A. A. J. Perenboom, P.Wyder, and F. Meier, Phys. Rep. 
{\bf 78}, 173 (1981). 

\bibitem{Nemeth90}R.~Nemeth, J. Stat. Phys., {\bf 63}, 419  (1991). 

\bibitem{Cheung94}H. P. Cheung, and H. F. Cheung, J. of Phys. Cond. Mat., 
{\bf 7}, 6707  (1995). 

\bibitem{foot1}Note that in the grandcanonical case, at $T=0$, there are 
no dynamical fluctuations due to an exchange of particles between the system 
and the reservoir. 



\bibitem{foot2}It should be stressed that equilibrium quantities, both within 
thermodynamics and statistical mechanics, are perfectly well defined even for 
small systems, as long as the reservoir involved is infinite. 


\bibitem{Rosenfeld61}R.~Rosenfeld, Physica, {\bf 27}, 67 (1961). 

\bibitem{Kubo75}M.~Toda, R.~Kubo, and N.~Sait\^{o}, {\em Statistical 
Physics}, Springer--Verlag, Berlin, 1983. 

\bibitem{foot3}
Note that $\Delta^{-1}=\partial \la N(\mu)\ra /\partial \mu\bigg|_{\bar\mu}=
-\partial^2\la \Omega(\mu)\ra /\partial \mu^2\bigg|_{\bar\mu}$. 

    
\bibitem{Negele}J. W. Negele, and H. Orland, {\em Quantum Many--Particle 
Systems}, Addison--Wesley, Redwood City, 1988. 


\bibitem{Kamenev95}A.~Kamenev,and  Y.~Gefen, 
    in  H.~A.~Cerdeira {\it et al.} (editors), 
    {\em Quantum Dynamics of Submicron Structures}, 
    Kluwer Academic Publ., Netherlands, pp. 81--92, 1995.
    


\bibitem{Altshuler86}B. L. Altshuler, and B. I. Shklovskii, Zh. Eksp. Teor.
Fiz. {\bf 91}, 220 (1986) [Sov. Phys. JETP {\bf 64}, 127 (1986)].


\bibitem{foot4}The 2--loop diagram of $w=2$ family (Eq.~(\ref{eq8})) is 
parametrically comparable to the 3--loop diagram of $w=1$ family (the latter 
comes with a small numerical value). Similar statements hold for higher--loop 
diagrams, which cause deviations of exact result from the one given by 
Eq.~(\ref{eq8}). Below we consider only the first two  terms of the 
expansion, which are given accurately by Eq.~(\ref{eq8}).

\bibitem{foot5}The coefficients are 
$\gamma_d=\pi^{-1}2^{-(3+d)}d(2-d)\int_0^\infty dx x^{1-d/2}
(x^{-2}-\mbox{sinh}^{-2}x)$, 
$\eta_d=\pi^{-3}2^{-(6+d/2)}(4-d)(2-d)\Gamma(4-d/2)\zeta(3-d/2)$. 

\bibitem{foot6}For a Poisson statistics the corrections to $C$ are small 
(proportional to the inverse band--width) but linear in $T$, in conjunction 
with this observation. 


\end{references}
\end{document}